\documentclass[aps,prl,groupedaddress,twocolumn]{revtex4}

\usepackage{enumerate}
\usepackage{amsfonts,amssymb,amsmath}
\usepackage[]{graphics,graphicx,epsfig}
\usepackage{amsthm}

\usepackage{graphicx}
\usepackage{dcolumn}
\usepackage{natbib}
\usepackage{color}
\usepackage{multirow}
\usepackage{ulem}

\begin{document}


\title{Galilean invariance without superluminal particles}


\author{Andrzej Grudka}
\affiliation{Institute of Spintronics and Quantum Information, Faculty of Physics, Adam Mickiewicz University, 61-614 Pozna\'n, Poland}

\author{Antoni W\'{o}jcik}
\affiliation{Institute of Spintronics and Quantum Information, Faculty of Physics, Adam Mickiewicz University, 61-614 Pozna\'n, Poland}


\date{\today}

\begin{abstract}
Recently Dragan and Ekert [New. J. Phys 22, 033038, 2020] presented arguments that probabilistic dynamics inherent in the realm of quantum physics is related to the propagation of superluminal particles. Moreover they argue that existence of such particles is a natural consequence of the principle of relativity. We show that the proposed extension of Lorentz transformation can be interpreted in natural way without invoking superluminal phenomena. 
\end{abstract}


\maketitle

\section{Introduction}
No particles (or signals) can move faster than light. Yet we sometimes use the notion of superluminal particles in order to better understand consequences of existence of the speed limit. For example Rindler [1, p. 54] considers superluminal signal going from spacetime point A to point B and causing some effect at that point (glass breaking). Rindler pointed out that there is another reference frame in which point B forerun point A. In this frame it seems that glass breaks spontaneously (without cause). Rindler concludes "Since in macro-physics no such uncaused events are observed, nature must have a way to prevent superluminal signals." Recently Dragan and Ekert [2] proposed to reconsider superluminal particles in the realm of quantum world in which intrinsic random events (which we cannot predict and which seems to have no causes) occurs. Although quantum mechanics theory is in full agreement with observations it is still true that in words of Dragan and Ekert "The notion of inherent randomness, something that happens without any cause, goes against our rational understanding of reality" [2]. These authors present arguments that non-deterministic dynamics is a natural consequence of the principle of relativity. Their result, if correct, would be very profound as connecting relativistic and quantum physics at very deep level. Their main argument is based on Galilean principle of relativity i.e. on postulated equivalence of all inertial frames without excluding the possibility of existence of superluminal ones. It is highlighted that this principle allows two branches of coordinate transformations (instead of single branch usually exploited in relativistic physics). Dragan and Ekert state that these two branches correspond to subluminal and superluminal families of observers. In this note we impair this correspondence.

\section{Transformations of coordinates}
It is sufficient for presenting our argument to consider simple $1+1$ dimensional case. For both time and space dimensions we will use coordinates $\xi_j (j=1,2)$ with the units of length. Let us consider two inertial observers $O$ and $O'$ with $O'$ moving relative to $O$ with dimensionless velocity $V$ (velocity is measured in the units of c) using coordinate systems $\xi_j$ and $\xi_j'$ respectively. Simple
arguments based on symmetry between $O$ and $O'$ (Fig. 1a, b) lead to linear transformation between coordinate systems in the usual form
\begin{eqnarray}
& \xi_1'=\gamma(V)(\xi_1-V\xi_2),\nonumber\\
& \xi_2'=\gamma(V)(\xi_2-V\xi_1)
\end{eqnarray}
with $\gamma(V)$ being symmetric or antisymmetric function of V.
Possible next step (in the common derivation of Lorentz transformation) is to consider $O$ and $O'$ with reversed space coordinates (see Fig. 1c).
Then transformation of Eqs. (1) gives
\begin{eqnarray}
& \tilde{\xi}_1'=\gamma(-V)(\tilde{\xi}_1-(-V)\tilde{\xi}_2),\nonumber\\
& \tilde{\xi}_2'=\gamma(-V)(\tilde{\xi}_2-(-V)\tilde{\xi}_1).
\end{eqnarray}
After using
\begin{eqnarray}
& \tilde{\xi}_1=\xi_1,\nonumber\\
& \tilde{\xi}_2=-\xi_2,\nonumber\\
& \tilde{\xi}_1'=\xi_1',\nonumber\\
& \tilde{\xi}_2'=-\xi_2'
\end{eqnarray}
in Eqs. (2) one obtains
\begin{eqnarray}
& \xi_1'=\gamma(-V)(\xi_1-V\xi_2),\nonumber\\
& \xi_2'=\gamma(-V)(\xi_2-V\xi_1),
\end{eqnarray}
and finally concludes that $\gamma(V)$ must be a symmetric function i.e. $\gamma(-V)=\gamma(V)$.

Dragan and Ekert propose to ignore this conclusion and to proceed further without fixing the parity of $\gamma(V)$. This is acceptable from the pure mathematical point of view. However, one has to be aware that in consequence standard (or natural) physical interpretation of mathematical expressions thus obtained can be misleading. Anyway let us proceed along the line proposed by Dragan and Ekert. They postulate that $\gamma(V)$ fulfils
\begin{eqnarray}
\frac{\gamma(V)\gamma(-V)-1}{V^2\gamma(V)\gamma(-V)}=K
\end{eqnarray}
with some constant $K$. (Note, than in the symmetric case $\gamma(-V)=\gamma(V)$, this postulate can be justified by considering three observers and mutual transformations between them (Eqs. (3) – (7) in [2].)

Now we can analyze two mentioned branches of transformation, which emerge after fixing the parity of $\gamma(V)$.

\begin{figure}
\includegraphics[width=8truecm]{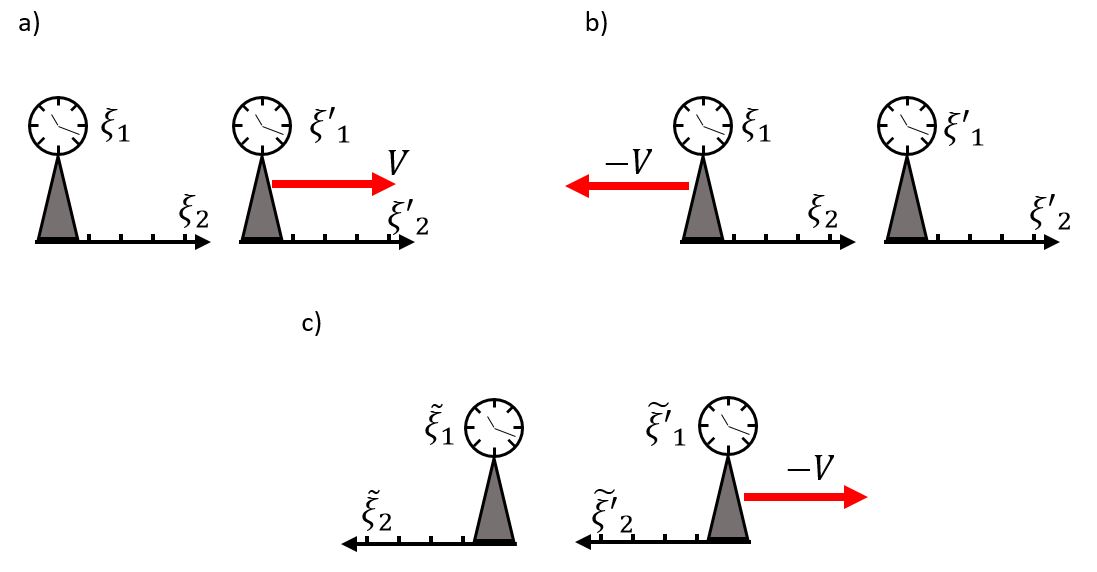}%
\caption{\label{Fig1}Three representations of observer $O'$ moving to the right with speed $V$ relative to the observer $O$.}
\end{figure}

\section{Symmetric case}
First let us consider the symmetric case $\gamma(-V)=\gamma(V)$. In this case Eq. (5) gives
\begin{eqnarray}
\gamma(V)=\pm\frac{1}{\sqrt{1-KV^2}}.
\end{eqnarray}
Thus we obtain a family (parametrized by $\tau, K, V$) of coordinate transformations in the form
\begin{eqnarray}
\begin{bmatrix}
\xi_1'\\
\xi_2'
\end{bmatrix}=\Lambda(\tau,K,V)\begin{bmatrix}
\xi_1\\
\xi_2
\end{bmatrix},
\end{eqnarray}
where
\begin{eqnarray}
\Lambda(\tau,K,V)=\tau \frac{1}{\sqrt{1-KV^2}}\begin{bmatrix}
1 & -V\\
-V&1
\end{bmatrix}
\end{eqnarray}
with $\tau=\pm1$ and $KV^2<1$.
Let us now emphasize that starting from concrete physical scenario (Fig. 1 and Eqs. (1)) we obtain continuum of possible coordinate transformations. This redundancy arises from the fact that observers can use arbitrary chosen coordinate systems. It is often highlighted (usually in the context of curved spacetimes) that coordinates have no direct physical meaning. Of course this is also true in flat spacetimes. Connection between coordinates and measurable quantities (distances and time intervals) has to be established with the use of some apparatus (e.g. clocks, meter sticks or teodolits, light signals) ([2], p. 41). This connection then can be represented by metric tensor. However, among many (abstract) coordinate systems one can choose coordinates established by physical means (grids of meter sticks and synchronised clocks). We will label such “measurement induced” coordinates by $ct$ and $x$. Metric tensor expressed in these coordinates has a simple form
\begin{eqnarray}
g=\begin{bmatrix}
1 & 0\\
0 & -1
\end{bmatrix}
\end{eqnarray}
and measurable space and time displacements are simply given by differences of coordinates. Note, that it is only after fixing both the physical scenario (Fig.1) and coordinate systems $\xi_1=c t$, $\xi_2=x$, $\xi_1'=c t'$, $\xi_2'=x'$ that one can fix parameters $\tau=1$, $K=1$ and interpret $V$ as velocity of $O'$ relative to $O$ restricted to $|V|\leq1$. In this case
\begin{eqnarray}
\begin{bmatrix}
c t'\\
x'
\end{bmatrix}=\Lambda(1,1,V)\begin{bmatrix}
c t\\
x
\end{bmatrix}.
\end{eqnarray}
One can as well use transformation $\Lambda(-1,1,V)$ with $\xi_1=c t$, $\xi_2=x$, $\xi_1'=-c t'$, $\xi_2'=-x'$ with the same interpretation of $V$. On the other hand if one decides to use e.g. $K=-1$ there will be no restrictions to $V$. In this case $|V|>1$ is of course not a sign of superluminality. It simply means that there is no possibility to interpret $V$ as a relative velocity between observers in this case. For example in the limit of infinite $V$
\begin{eqnarray}
\Lambda(1,-1,\infty)=\begin{bmatrix}
0 & -1\\
-1 & 0
\end{bmatrix},
\end{eqnarray}
which can be interpreted as transformations between two observers at rest.

\section{Antisymmetric case}
Let us now turn to the antisymmetric case. To make clear distinction between two branches of transformation (originated from the choice of the $\gamma(V)$ parity) we introduce some new notation in the antisymmetric case. Namely parameter $V$ will be denoted by W and coordinates used by observer $O'$ will be denoted by $\eta_j$. $\gamma(-W)=-\gamma(W)$ leads to
\begin{eqnarray}
\gamma(W)=\pm\frac{W}{|W|}~\frac{1}{\sqrt{KW^2-1}}.
\end{eqnarray}
It follows that
\begin{eqnarray}
\begin{bmatrix}
\eta_1\\
\eta_2
\end{bmatrix}=L(\tau,K,W)\begin{bmatrix}
\xi_1\\
\xi_2
\end{bmatrix}
\end{eqnarray}
with
\begin{eqnarray}
L(\tau,K,W)=\tau~ \frac{W}{|W|}~\frac{1}{\sqrt{KW^2-1}}\begin{bmatrix}
1 & -W\\
-W&1
\end{bmatrix}.
\end{eqnarray}
In this case it is not so obvious from the physical point of view, how to choose the value of $K$. For simplicity let us choose $K=1$ and concentrate on physical interpretation of the transformation $L(-1,1,W)$ with $|W|\geq1$.
In Fig. 2 we present spacetime diagrams showing worldlines of some massive (with non-zero invariant mass) subluminal (from the point of view of observer $O$) particles and light rays. First notice that light rays seem to propagate in expected way. In both coordinate systems light rays fulfil the same equation $\xi_1=\pm \xi_2$ and $\eta_1=\pm \eta_2$. The most important observation,
however, is that looking at Fig. 2b it seems that particles are propagating with
superluminal velocities. Is this observation correct?

\begin{figure}
\includegraphics[width=8truecm]{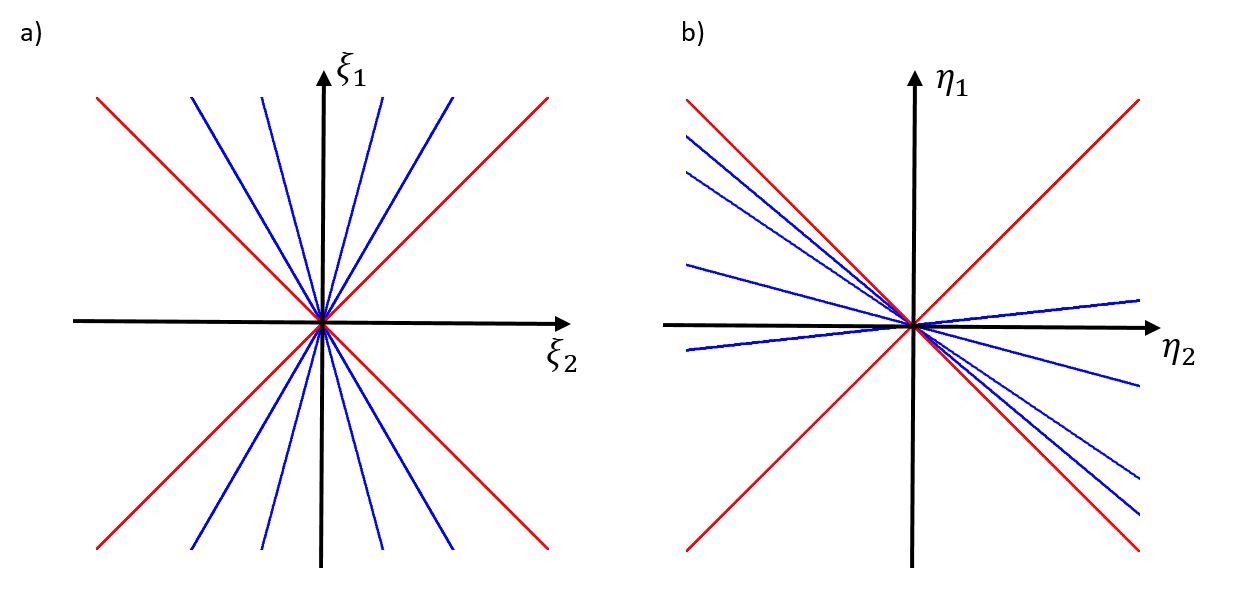}%
\caption{\label{Fig2}Blue lines are the worldlines of four particles (subluminal from the point of
view of $O$ observer) and red are light rays. Transformation between coordinates
systems is given by $L(-1,1, 2)$.}
\end{figure}

Before answering this question, let us notice that this observation is crucial for
the argument of Dragan and Ekert. Because, if we have a given particle which
propagates subluminally and superluminally with respect to two equally "legal"
observers, than considering superluminal particles seriously cannot be avoided.
On the other hand the existence of superluminal particles cannot be compatible
with local and deterministic dynamics, hence unexpected and exciting
connection between relativity postulate and inherent randomness of quantum
mechanics can be obtained.

However, we suggest that the above observation could be wrong.
Observation, that particles in Fig. 2b propagate superluminally is based on
hidden assumption that there exist observer $O'$ (called by Dragan and Ekert
superluminal observer) whose measurements of time interval $\Delta t_{O'}$ and distance
$\Delta x_{O'}$ are connected to coordinates $\eta_j$ in a simple way
\begin{eqnarray}
& c \Delta t_{O'}=\Delta \eta_1,\nonumber\\
& \Delta x_{O'}=\Delta \eta_2.
\end{eqnarray}
It seems, however, that this assumption is based just on the analogy between
Eqs. (10) and Eqs. (13) and not on any actual consistent construction of
"measurement induced" coordinate system for $O'$.

Let us show
that one can choose more natural interpretation of $\eta_j$ coordinate system. Fig. 3a
presents space-time diagram with two light rays emitted from the origin of
coordinate system to the left and to the right (from the point of view of observer
$O$). Fig. 3b presents the same rays in the $\eta_j$ coordinates diagram. What we mean
by natural interpretation is to assume that the horizontal axis in Fig. 3b
represents time coordinate, whereas vertical one represents space coordinate.
It means that transformation $L(-1,1, W)$ interchanges space and time
coordinates. We have seen similar effect  before in transformation $\Lambda(1,1, \infty)$.
Traversing Schwarzschild horizon also results in exchange of time and radial coordinates. So we propose to
replace Eqs. (15) by
\begin{eqnarray}
& c \Delta t_{O'}=\Delta \eta_2,\nonumber\\
& \Delta x_{O'}=\Delta \eta_1.
\end{eqnarray}
With this choice of coordinates particles in Fig. 2b propagate of course with subluminal velocity.
We can now prove that one can construct measurement system consistent with our proposition in Eqs. (16) such that these equations together with Eq. (13) lead to
\begin{eqnarray}
\begin{bmatrix}
c \Delta t_{O'}\\
\Delta x_{O'}
\end{bmatrix}=\begin{bmatrix}
0 & 1\\
1 & 0
\end{bmatrix}L(-1,1,W)\begin{bmatrix}
c \Delta t_{O}\\
\Delta x_{O}
\end{bmatrix}.
\end{eqnarray}
It is easy to show that
\begin{eqnarray}
L(-1,1,W)=\begin{bmatrix}
0 & 1\\
1 & 0
\end{bmatrix}\Lambda(1,1,W^{-1}).
\end{eqnarray}
Thus coordinate transformation $L(-1,1,W)$ with parameter $W\geq1$ together with interchanging space and time coordinates is equivalent to standard transformation $\Lambda(1,1,V)$ with velocity $V=W^{-1}\leq1$. It follows that standard textbooks construction based on meter sticks and clocks is consistent with Eqs. (16) (see also Fig. 4). Moreover we have
\begin{eqnarray}
& L^{-1}(-1,1,W)=\Lambda(1,1,-W^{-1})
\begin{bmatrix}
0 & 1\\
1 & 0
\end{bmatrix}=\nonumber\\
& =L(-1,1,-W).
\end{eqnarray}
It should be so because in order to return to original coordinate system we should first exchange space and time coordinates and then perform Lorentz transformation with opposite velocity. This leads to antisymmetric $\gamma(W)$ as given in Eq. (12).
\begin{figure}
\includegraphics[width=8truecm]{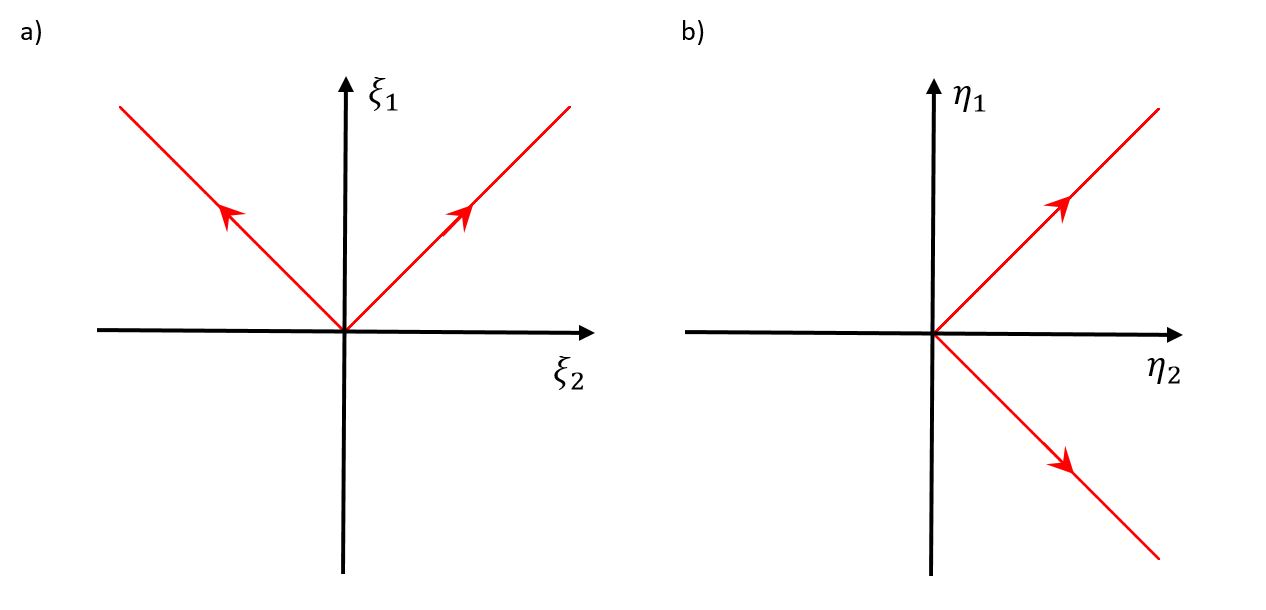}%
\caption{\label{Fig3.}Two light rays emitted from the common origin of the both coordinate
systems. Transformation between coordinates systems is given by $L(-1,1, 2)$.}
\end{figure}

\begin{figure}
\includegraphics[width=4truecm]{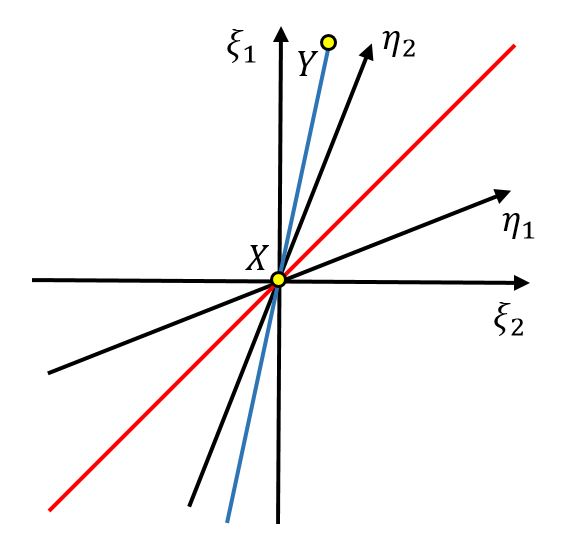}%
\caption{\label{Fig4.}Minkowski diagram for observers $O$ and $O'$ with light ray (red) and particle worldline (blue). $X$ and $Y$ are two events on particle worldline.}
\end{figure}

Let us emphasize that coordinate system $(\eta_1,\eta_2)$ can only be called superluminal because equation of motion of the point $\eta_2=0$ as seen by observer $O$ is
\begin{eqnarray}
    x=Wc~t,
\end{eqnarray}
with $W>1$. However, it is well known that "Arbitrary large velocities are possible for moving points that carry no information (...)" ([1], p. 56).

\section{Galilean invariance and (non)relativity of superluminality}
Let us point that there are two possible notions of superluminality. First one (let us call it coordinate superluminality) is based on coordinate velocity. Second one (geometric superluminality) considers spacetime displacement as superluminal if the square of the spacetime interval is  less than $0$ (${\Delta s}^2<0$ ). For standard subluminal coordinate systems these two notions are equivalent. However, in the case of superluminal coordinate systems they disagree. In Fig. 6 particle wordline connecting two events - $X$ and $Y$  is presented. It is clear that 
\begin{eqnarray}
    \left|\frac{\Delta\xi_2}{\Delta\xi_1}\right|<1
\end{eqnarray}
whereas
\begin{eqnarray}
    \left|\frac{\Delta\eta_2}{\Delta\eta_1}\right|>1.
\end{eqnarray}
Thus, according to the first notion of superluminality the particle is subluminal relative to observer O and superluminal relative to $O'$. 
\newline\indent On the other hand let us consider square of the spacetime interval between events $X$ and $Y$
\begin{eqnarray}
    {\Delta s}^2=(\Delta\xi_1)^2-(\Delta\xi_2)^2.
\end{eqnarray}
Although expressed here with the use of coordinates, this is purely geometrical notion, i.e. has absolute, independent on the coordinates meaning. It can be stated e.g. as a proper time measured by clock moving from $X$ to $Y$ with constant speed (see [1], p. 93).  All observers (independent on their motion and on coordinates systems which they use) have to agree about (so measured) proper time. From Fig. 4 or Eq. (23) one sees that ${\Delta s}^2>0$. So it must be that ${\Delta s}^2>0$ for observer $O'$ too. Of course one can as well express ${\Delta s}^2$ with the use of ($\eta_1$,$\eta_2$) coordinates. Using Eq. (13) in Eq. (23) gives 
\begin{eqnarray}
    {\Delta s}^2=-(\Delta\eta_1)^2+(\Delta\eta_2)^2.
\end{eqnarray}
Once again it can be seen that $\eta_1$ plays role of spatial coordinate, whereas $\eta_2$ plays role of time coordinate. 
\newline \indent So (in the sense of geometric superluminality) the particle wordline depicted in Fig. 4 must be considered as subluminal for both observers $O$ and $O'$.
\newline \indent Note that what is needed in the argument of Dragan and Ekert is the geometric superluminality (possibility of sending particles between spatially separated events ${\Delta s}^2<0$). Unfortunately, what is provided by introducing second branch of transformations is only the existence of coordinate superluminality. Causal partition of events by light cones is absolute, not only for users of Lorentz transformations $\Lambda\left(1,1,V\right)$, but also for users of newly introduced transformations $L\left(-1,1,W\right)$. It is not true that particle at rest relative to observer $O$ will be considered superluminal by observer $O'$ (in the geometric sense of superluminality). Being superluminal is not relative.
\section{Conclusions}
We prove that the second branch of coordinate transformation (based on  antisymmetricity of $\gamma\left(V\right))$ given by Eq. (14) can be interpreted in consistent way without notion of superluminal particles or superluminal observers. It does not mean that we prove that such superluminal phenomena cannot exist. It means only, that introducing some new coordinates is not enough to prove existence of them. It must be remembered that coordinates alone have no physical meaning. Moreover, we show that proposed second branch of transformations is coherent with standard causal structure induced by light cones and expressed by the sign of spacetime interval squared.

\begin{acknowledgments}
We thank A. Dragan for comment on the manuscript. This work has been supported by the Polish National Science Centre (NCN) under the Maestro Grant No. DEC-2019/34/A/ST2/00081.
\end{acknowledgments}

\bibliography{basename of .bib file}

\end{document}